\documentclass[conference]{IEEEtran}
\IEEEoverridecommandlockouts
\usepackage{enumitem}
\usepackage{cite}
\usepackage{amsmath,amssymb,amsfonts}
\usepackage{algorithmic}
\usepackage{enumitem}
\usepackage{graphicx}
\usepackage{textcomp}
\usepackage{xcolor}
\usepackage{csquotes}
\usepackage{balance}
\usepackage{url}
\usepackage{balance}
\usepackage{graphicx}
\usepackage{epstopdf}
\def\BibTeX{{\rm B\kern-.05em{\sc i\kern-.025em b}\kern-.08em
    T\kern-.1667em\lower.7ex\hbox{E}\kern-.125emX}}

\usepackage{soul, color}
\usepackage[acronym,toc,shortcuts]{glossaries}
\usepackage{multirow}

\graphicspath{{figures/}}

\makeglossaries
\newacronym{3GPP}{3GPP}{The 3rd Generation Partnership Project }
\newacronym{5G}{5G}{Fifth Generation}
\newacronym{AAA}{AAA}{Authentication, Authorization and Accounting}
\newacronym{ARIMA}{ARIMA}{AutoRegressive Integrated Moving Average}
\newacronym{CPU}{CPU}{Central Processing Unit}
\newacronym{GRU}{GRU}{Gated Recurrent Unit}
\newacronym{O-RAN}{O-RAN}{Open Radio Access Network}
\newacronym{QoS}{QoS}{Quality of Service}
\newacronym{PRB}{PRB}{Physical Resource Block}
\newacronym{rApp}{rApp}{radio App}
\newacronym{RAN}{RAN}{Radio Access Network}
\newacronym{RIC}{RIC}{RAN Intelligent Controller}
\newacronym{xApp}{xApp}{eXtended application}
\newacronym{SFF}{SFF}{Simple-Feed-Forward}
\newacronym{LSTM}{LSTM}{Long-Short Term Memory}
\newacronym{SN}{SN}{Seasonal-Naive}
\newacronym{MLP}{MLP}{ Multi-Layer Perceptron}
\newacronym{RNN}{RNN}{Recurrent Neural Network}
\newacronym{MSE}{MSE}{Mean Square Error}
\newacronym{MAE}{MAE}{Mean Absolute Scaled Error}
\newacronym{MAPE}{MAPE}{Mean Absolute Percentage Error}
\newacronym{ND}{ND}{Normalized Deviation}
\newacronym{QL}{QL}{Quantile Loss}
\newacronym{AI}{AI}{Artificial Intelligence}
\newacronym{ML}{ML}{Machine Learning}
\newacronym{SLAs}{SLAs}{Service Level Agreements}
\newacronym{NN}{NN}{Neural Network}
\newacronym{BS}{BS}{Base Station}

\usepackage{fancyhdr}
\usepackage{geometry}
\usepackage{lipsum} 

\begin{document}

\title{ On the Impact of PRB Load Uncertainty Forecasting for Sustainable Open RAN
}


\author{Vaishnavi Kasuluru, Luis Blanco, Cristian J. Vaca-Rubio, Engin Zeydan \\
{\normalsize{} Centre Tecnològic de Telecomunicacions de Catalunya (CTTC/CERCA), Castelldefels, Barcelona, Spain, 08860.} \\
{\normalsize{} Emails: \texttt{\{vkasuluru, lblanco, cvaca, ezeydan\}@cttc.es}}

\thanks{This work has been supported by SEMANTIC project, funded by the European Union’s Horizon 2020 research and innovation program under the Marie Skłodowska-Curie grant (agreement No 861165), the Horizon Europe project VERGE (ID: 101096034), Spanish MINECO - Program UNICO I+D (grants TSI-063000-2021-54 and -55) funded by MINECO through the “NextGenerationEU” program, and the Spanish project ORIGIN (PID2020-113832RB-C22) funded MICCIN}


}

\maketitle

\begin{abstract}

The transition to sustainable \ac{O-RAN} architectures brings new challenges for resource management, especially in predicting the utilization of \ac{PRB}s. In this paper, we propose a novel approach to characterize the \ac{PRB} load using probabilistic forecasting techniques. First, we provide background information on the O-RAN architecture and components and emphasize the importance of energy/power consumption models for sustainable implementations. The problem statement highlights the need for accurate \ac{PRB} load prediction to optimize resource allocation and power efficiency. We then investigate probabilistic forecasting techniques, including \ac{SFF}, DeepAR, and Transformers, and discuss their likelihood model assumptions. The simulation results show that DeepAR estimators predict the \ac{PRB}s with less uncertainty and effectively capture the temporal dependencies in the dataset compared to \ac{SFF}- and Transformer-based models, leading to power savings. Different percentile selections can also increase power savings, but at the cost of over-/under provisioning. At the same time, the performance of the \ac{LSTM} is shown to be inferior to the probabilistic estimators with respect to all error metrics. Finally, we outline the importance of probabilistic, prediction-based characterization for sustainable O-RAN implementations and highlight avenues for future research.



\end{abstract}



\section{Introduction}
\label{intro}

Recent forecasts suggest that by 2030, Information and Communication Technology (ICT) networks could consume up to 21\% of the world's electricity supply \cite{enerdata_report}. In addition, the entire ICT sector contributes over 2\% to global greenhouse gas emissions \cite{freitag2021real}. To put this into perspective, this level of emissions corresponds to that of the aviation industry as a whole. This projection raises concerns about the environmental, economic and social sustainability of these networks. Consequently, the integration of sustainability principles and power efficiency becomes essential in the development and deployment of 6G networks.   Having a closer look at the recent history of cellular networks, although the deployed 5G networks are roughly four times more energy-efficient compared to their 4G networks, their energy consumption is approximately three times greater \cite{piovesan2023power}. This is primarily due to the need for a greater number of cells to maintain equivalent coverage at higher frequencies, as well as the higher processing requirements resulting from wider bandwidths and a greater number of antennas. In this context, the use of energy is crucial in terms of sustainable operation. In mobile networks, energy consumption is a major concern for operators, as it significantly increases electricity bills and thus affects their Operational Expenditure (OPEX). To minimize energy consumption in mobile networks, it is important to understand where energy is consumed within the network. The \ac{RAN} is responsible for a significant part of the energy consumption in mobile networks, with the O-RU component accounting for the largest share ( around 60\% of the total energy of a base station). An examination of the breakdown of energy consumption within the mobile network shows that the \ac{RAN} accounts for 73\% of the total energy consumption, followed by the core network with 13\%, the data centers with 9\%, and other operational aspects with 5\% \cite{NGM_2021} \cite{GSMA_2021}. 

Pioneer studies, such as the paper in reference \cite{auer2011much}, represent one of the most widely used \ac{BS} power consumption models in the literature and show the linear relationship between total BS power consumption and the transmit power. A recent work in \cite{piovesan2022machine} provided a realistic characterization of 5G multi-carrier \ac{BS}s, giving an analytical energy consumption model based on large data collection campaigns. The works in \cite{piovesan2023power} and \cite{piovesan2022machine} showed that in modern BSs, the power consumption increases linearly with the \ac{PRB} load. Furthermore, as has been recently shown in \cite{lopez2023data}, DL \ac{PRB}
load, i.e., the ratio between the used \ac{PRB}s in a BS and the maximum number of \ac{PRB}s available at the remote unit, holds the highest significance in modeling radio unit energy consumption. In general, BSs are dimensioned to serve a large amount of traffic during busy hours and in practice this leads to high underutilized bandwidth usage during the major part of the day. Monitoring and managing DL \ac{PRB} load is of paramount importance for optimizing network performance and ensuring quality of service for users. It involves dynamically allocating resources and optimizing network configurations to accommodate varying traffic demands and maintain energy efficient operation.

\ac{O-RAN} is a new communication paradigm designed to enable the next generation of communication systems. It provides a transformative architectural approach to mobile networks that emphasizes openness, interoperability, and innovation within the radio access network ecosystem. \ac{O-RAN} promotes the separation of network elements and the introduction of open interfaces, leading to greater choice of providers and greater flexibility in deployment. In the open RAN domain, the concept of radio applications (rApps) is fundamental as it extends the capabilities of the \ac{RIC} and enables advanced features such as AI-enhanced proactive allocation of radio resources. In the O-RAN architecture, this paper proposes the integration of AI-enhanced probabilistic forecasting  models as an rApp tool to predict the \ac{PRB} requests. Compared to conventional single-point time series forecasting techniques (e.g. \ac{LSTM}s \cite{hochreiter1997long} or \ac{GRU}s \cite{cho2014learning}), state-of-the-art (SotA) probabilistic forecasting techniques (e.g. DeepAR \cite{salinas2020deepar}, Transformers \cite{vaswani2017attention}) are able to quantify the uncertainty in the prediction, which allows for more informative and reliable decisions. This paper investigates the power saving and error performance of the forecasting models \ac{SFF}, DeepAR, and Transformer and compares them with the deterministic single-point estimator LSTM. Our results show that DeepAR estimators can predict the \ac{PRB}s with lower uncertainty and effectively capture the temporal dependencies in the dataset compared to \ac{SFF}- and Transformer-based models, leading to power savings. Different percentile selections in decision engine for probabilistic methods can also increase power savings, but at the cost of over-/under provisioning. At the same time, the performance of the \ac{LSTM} is shown to be inferior to the probabilistic estimators with respect to all error metrics. 

The rest of the paper is organized as follows. Section \ref{background} provides background information on the O-RAN architecture, its components and gives a power consumption analytical model. Section \ref{problem} provides the problem statement. Section \ref{probabilistic} gives an overview the probabilistic forecasting methods, namely \ac{SFF}, DeepAR, and Transformer. Section \ref{simulations} presents the simulation results and finally Section \ref{conclusions} provides the conclusions and future direction of the paper.

\par

\section{Background Information}
\label{background}

\subsection{\ac{O-RAN} Architecture and Components}
 
\begin{figure*}[htp!]
\centering
\includegraphics[width=.9\linewidth]{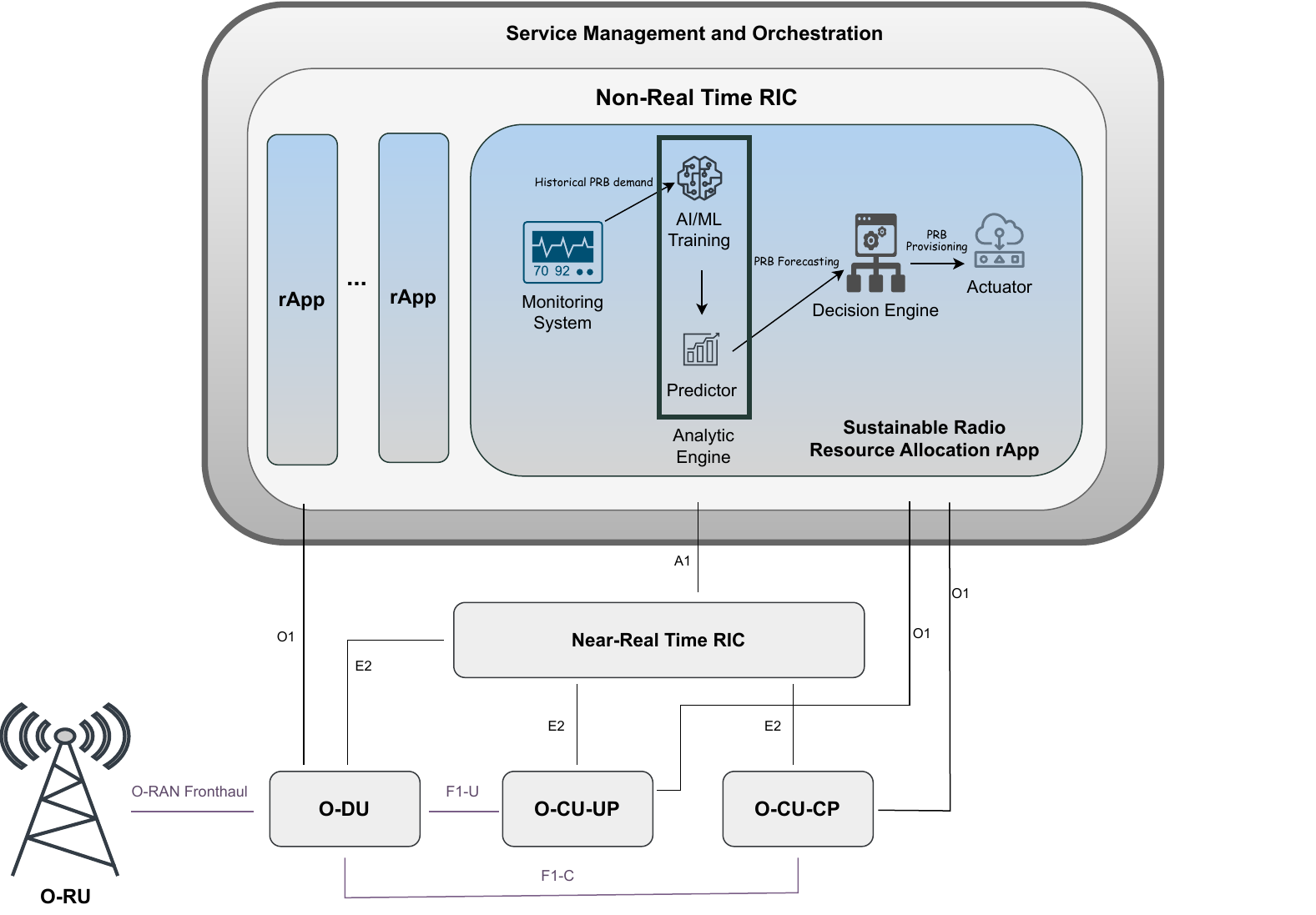}
\caption{\ac{O-RAN} architecture with sustainable radio resource allocation as \ac{rApp}.}
\label{O-RAN}
\end{figure*}

Traditional \ac{RAN} components, up to the 4th generation, are mostly hardware-dependent. They were highly vendor-dependent, which made the integration of a cooperative intelligent network very complicated. Updating hardware components or proprietary systems significantly increases CAPEX and OPEX costs. Further development of RAN solutions for the next generation is proposed as a solution to these problems. Virtualization and disaggregation are the basis of O-RAN technology. O-RAN aims to enable open and intelligent resource management with universally compatible software solutions and minimalist hardware to avoid vendor lock-in. The most important aspects of O-RAN include open interfaces, centralised orchestration and interoperability. The most important key elements are the Open-Radio Unit (O-RU), the Open-Distributed Unit (O-DU), the Open-Central Unit (O-CU), the Near-Real-Time \ac{RIC}, and the Non-Real-Time \ac{RIC}, as shown in Fig.\ref{O-RAN}. 

The functions of the Open-Radio Unit (O-RU), the Open-Distributed Unit (O-DU) and the Open-Central Unit (O-CU) are similar to those of the disaggregated 5G-RAN, but with additional support for \ac{O-RAN} specifications and interfaces. The near-real-time \ac{RIC} helps to optimize resources and control RAN elements based on fine-grained data sets with AI/ML-based applications. It is suitable for tasks with a low latency overhead of 10ms to 1s. Non-real-time \ac{RIC} controls and optimizes the resource based on coarse-grained broad datasets for applications with latency requirements of more than 1s. It also helps to provide policy-based guidance to near-real-time \ac{RIC}. The sustainable radio resource allocation strategy presented in this paper is considered as rApp in non-real-time RIC. This rApp consists of 4 main components, namely:

\begin{itemize}[leftmargin=2.8mm]
    \item \textbf{Monitoring System} gathers the historical data from the O-DU about assigned DL \ac{PRB}s and forwards it to the analytic engine and other elements that request the data.
    \item \textbf{Analytic Engine} pre-processes the data and splits it into a training and a test set. During training and prediction phases, the data is passed as an input feature to the various probabilistic forecasting estimators for analysis and prediction of the future PRB demands. Probabilistic forecasting techniques predict a range of possible values as well as their associated probabilities, enabling a more realistic representation of future events.
    \item \textbf{Decision Engine} receives as an input the estimation of \ac{PRB}s with their uncertainty from the analytic engine. As introduced in the next section, there is trade-off between fulfilling the PRB demands and power consumption. The decision engine must choose the PRBs to be provisioning based on the sensitivity to over-/under-estimation, taking into account their impact on sustainability.  
    
    \item \textbf{Actuator} is the entity responsible for executing the actual PRB allocation in the dis-aggregated RAN.
\end{itemize}

\subsection{Power Consumption Model}
Let us consider the power consumption model for 5G BS introduced in \cite{piovesan2022machine} and \cite{song2023high}. The total power consumption, denoted by  $P$, can be mathematically formulated as

\begin{equation}
    P = P_0 + P_{BB} + P_{Tran} + P_{PA}+ P_{out}, \label{eq:power}
\end{equation} 
where $P_0$ denotes the baseline power consumption in sleep mode, $P_{BB}$ is the baseband processing power consumption. $P_{Tran}$ denotes the power consumption by the RF chains, the Power Amplifier consumption is $P_{PA}$, and $P_{out}$ is the power required for data transmission.

\begin{equation}
   P_{out} = \frac{1}{\eta}P_{\text{TX}} \frac{R_{BS}}{C_{BS}}, \label{eq:P_out}
\end{equation}

The first terms, i.e., $P_0$, $P_{BB}$,$P_{Tran}$, and $P_{PA}$, depend on the number of available and active RF chains. Following the approach in \cite{song2023high}, for the sake of simplicity, the first four terms in  \eqref{eq:power} are assumed as known and the fifth term is proportional to the traffic volume.

where $\eta$ denotes the efficiency of the power amplifiers and $P_{\text{TX}}$ is the maximum transmit power. $R_{BS}$ and ${C_{BS}}$ denote the actual rate and the capacity of the BS. The total capacity of the BS can be computed using the classical Shannon-Harley theorem and is given by

\begin{equation}
   C = B \log_2 \left(1 + \text{SINR}\right).
\end{equation}

Considering the capacity formula is clear that $P_{out}$ in  \eqref{eq:P_out} is inversely proportional to the total bandwidth in the O-RU. 
A natural surrogate of $\frac{R_{BS}}{C_{BS}}$ in (\ref{eq:P_out}) is to consider the DL \ac{PRB} load, expressed as the ratio between the average number of DL \ac{PRB}s and the maximum number of DL \ac{PRB}s available. It has been recently shown in \cite{lopez2023data} and \cite{piovesan2023power} that DL \ac{PRB} load holds the highest significance in
modeling radio unit power consumption. Furthermore, note that the transmit power increases linearly with the number of used \ac{PRB}s \cite{piovesan2022machine}. Therefore, there is a trade-off between satisfying the DL \ac{PRB} demands and the energy consumption.

\section{Problem Statement}
\label{problem}

We formulate the \ac{PRB} allocation forecasting problem as a time series at time $t$ by $y_{t}$, then our goal is to model the conditional distribution 
\begin{equation}
    P(\mathbf{y}_{t_0:T}|\mathbf{y}_{1:t_0-1})\label{eq:pf}
\end{equation} 
of the future of each PRB allocation value in the time series $[y_{t_0}, y_{t_0+1}, ..., y_{T}] := \mathbf{y}_{t_0:T}$ given its past $[y_{1}, ..., y_{t_0-2}, y_{t_0-1}] := \mathbf{y}_{1:t_0-1}$, where $t_0$ denotes the time point from which we assume $y_{t}$ to be unknown at prediction time. To avoid confusion, we refer to time ranges $[1, t_0-1]$ and $[t_0, T]$ as the conditioning and prediction ranges, respectively. Once we learn to forecast and quantify the uncertainty in the prediction range, we provide a sustainability analysis using eq (\ref{eq:P_out}).

\subsection{Probabilistic Forecasting Techniques}
\label{probabilistic}

In the field of time series forecasting, accurate predictions are mandatory for effective decision-making across various domains. While traditional forecasting methods offer deterministic point estimates, the characterization of the uncertainty in predictions provides valuable information for a proper assessment of decisions in open RAN networks. Due to recent advancements, deep learning algorithms are being integrated with traditional methods. Deterministic classical \ac{AI} forecast models like \ac{LSTM}, \ac{GRU}s, etc., fail to provide certainty about future forecasts.  Their overconfidence in forecasts emerges from ignorance of data uncertainty.

Among the different techniques available in the literature, three architectures have gained considerable attention: the \ac{SFF}, and DeepAR \cite{salinas2020deepar}, and Transformer \cite{vaswani2017attention}. They will deliver more accurate and representative predictions in the form of probability distributions. We here provide a comprehensive explanation of the models evaluated in this work.



\subsubsection{Simple FeedForward (SFF)}
SFF is based on a simple feed-forward \ac{NN} that estimates the probabilistic distribution of the allocated PRBs along the time series. Instead of doing single point predictions, the \ac{NN} will output the parameters of a desired distribution for every time step $t$. The network is composed of an input layer with neurons equal to the number of time steps in the conditioning range $[1, t_0-1]$, a hidden layer $h$, and an output layer with the number of neurons equal to the number of time steps in the prediction range $[t_0, T]\times p$, where $p$ denote the amount of parameters of the assumed likelihood model.
The output layers estimate the parameters of a probability distribution for each $t$ representing the forecast uncertainty. In our work, we assume this likelihood to be the t-student location-scale distribution given by:
\begin{equation}
    l(y_t|\nu,\mu,\sigma^2) = \frac{\Gamma\left(\frac{\nu+1}{2}\right)}{\sigma\sqrt{\nu\pi}\,\Gamma\left(\frac{\nu}{2}\right)} \left(\frac{\nu +(\frac{y_t-\mu}{\sigma})^2}{\nu}\right)^{-\frac{\nu+1}{2}},
\end{equation}
where $\mu(\mathbf{h}_t) = \mathbf{W}^T_\mu \mathbf{h}_t + \mathbf{b}_\mu$, $\sigma(\mathbf{h}_t) = \log (1 + exp(\mathbf{W}^T_\sigma \mathbf{h}_t + \mathbf{b}_\sigma))$ and $\nu(\mathbf{h}_t) = \log (1 + exp(\mathbf{W}^T_\nu \mathbf{h}_t + \mathbf{b}_\nu))$, and $\mathbf{h}_t$ denotes the output of the last layer at prediction time $t$. In this way, the $\mu$ is characterized directly by the network output, and $\sigma$ and $\nu$ are obtained by applying an affine transformation followed by a softplus activation to ensure $\sigma > 0$ and $\nu > 0$.
\subsubsection{DeepAR}
DeepAR is a probabilistic forecasting algorithm based on \ac{RNN} equipped with \ac{LSTM} units. Unlike traditional forecasting methods, deepAR generates probabilistic forecasts providing probability distributions over future \ac{PRB} allocation along the time series for every $t$ in the prediction range. In this case, to approximate eq (\ref{eq:pf}) we assume that the model distribution:
\begin{equation}
    \mathcal{Q}_{\Theta}(\mathbf{y}_{t_0:T}|\mathbf{y}_{1:t_0-1}),
\end{equation}
consist of a product of likelihood factors:
\begin{equation}
\begin{split}
    & \mathcal{Q}_{\Theta}(\mathbf{y}_{t_0:T}|\mathbf{y}_{1:t_0-1}) \\
    &= \prod_{t=t_0}^T \mathcal{Q}_{\Theta}(\mathbf{y}_{t_0:T}|\mathbf{y}_{1:t_0-1}) = \prod_{t=t_0}^T l(y_t|\theta_d(\mathbf{h}_t))
\end{split}
\end{equation}
which is parametrized by the output $\mathbf{h}_t$ of an autoregressive reccurent network $\mathbf{h}_t=h(\mathbf{h}_t, y_{t-1}, \Theta)$. In this case, $h$ denotes a function implemented by a multilayer \ac{RNN} with \ac{LSTM} cells.
\begin{equation}
    l(y_t | \mu, \sigma) = \frac{1}{\sigma\sqrt{2\pi}} e^{-\frac{(y_t - \mu)^2}{2\sigma^2}}
\end{equation}
Similarly to \ac{SFF} the model outputs the parameters of a probability distribution for each $t$. For DeepAR we assume a gaussian likelihood given by equation 8, where $\mu(\mathbf{h}_t) = \mathbf{W}^T_\mu \mathbf{h}_t + \mathbf{b}_\mu$ and $\sigma(\mathbf{h}_t) = \log (1 + exp(\mathbf{W}^T_\sigma \mathbf{h}_t + \mathbf{b}_\sigma))$.  Similarly to \ac{SFF}, $\sigma$ is obtained by applying an affine transformation followed by a softplus activation to ensure $\sigma > 0$.

\subsubsection{Transformer}
Transformers have emerged as a powerful architecture due to their ability to capture long-range dependencies in sequences. The transformer architecture comprises two main components: the encoder and the decoder. The encoder consists of a series of $N$ blocks, each comprising a Multi-Head Self-Attention layer followed by a position-wise fully connected layer with ReLU activations. These blocks process the input \ac{PRB} allocation time series in parallel, obtaining an encoded representation of this \ac{PRB}s. Next,
the decoder has three layers. The first and the last one are similar to the encoder and the second one is an encoder-decoder attention mechanism\footnote{In the interest of space, we refer the reader to \cite{vaswani2017attention} for a more detailed explanation.}. Similarly to \ac{SFF}, we assume t-student location-scale distribution, eq(5).
\subsubsection{Training loss}
Given the \ac{PRB} allocation time series $\mathbf{y}_{1:T}$, all our presented methods use the same loss for learning the parameters of the networks, summarized for simplicity as $\theta=\{\theta_s, \theta_d, \theta_t\}$. This is done by maximizing the log-likelihood:
\begin{equation}
    \mathcal{L}=\sum_{t=t_0}^T\log l(y_t|\theta(\mathbf{h}_t)).
\end{equation}
Furthermore, the models learn to estimate the distribution parameters for every time step $t$ in the prediction range.

\section{Simulation Results}
\label{simulations}
This section shows the analysis of the performance of different probabilistic estimators together with the deterministic \ac{LSTM} model. Python programming was used together with the Gluonts library to develop and analyze the estimators in the form of rApp. Herein, the 3 main estimators used are \ac{SFF}, DeepAR, and Transformer, which predict the DL \ac{PRB}s needed for the next 24-hour period based on 10 weeks of historical data. For \ac{SFF} estimator the hyperparameters considered are: epochs=5, batch size=1, hidden layer dimension=[40,40], and number of evaluation samples=100.
For DeepAR, the setup hyperparameters are the following: epochs=5, batch size=1, \ac{RNN} Layers=2, number of cells per \ac{RNN}=40, and number of evaluation samples=100. Finally, in the case of transformer the used hyperparameters are: epochs=5, batch size=1,  number of evaluation samples=100, dimension of transformer network=32, inner-hidden layers of transformer's feedforward network dimension=4; and context length=24. For analysis of the effect of \ac{PRB}s on power saving, we consider a single carrier and 64 RF chains. We are also considering the normalized values for the fitted parameters $P_0, P_{BB}, P_{Tran}, P_{PA}, P_{TX}, \eta $ in equations \eqref{eq:power} and \eqref{eq:P_out} as 0.22, 0.16, 0.09408, 0.24382, 43dBm, and 0.4, respectively as in paper \cite{lopez2023data}. The maximum available \ac{PRB}s for base station is 160, which corresponds to a 30MHz bandwidth. Here, the power saving computation is based on parameter $P_{out}$.

The historical \ac{PRB} data can be collected in the O-RAN architecture via the O1 interface from the O-DU. During pre-processing in the analytic engine, the DL \ac{PRB} data is first split into training and test data in a ratio of 80:20. The training data is then forwarded to the estimators as an input feature. Later, the test data is used to forecast, evaluate and compare the performance of the models in the prediction phase.

\begin{figure}[htp!]
\centering
\includegraphics[width=1\linewidth]{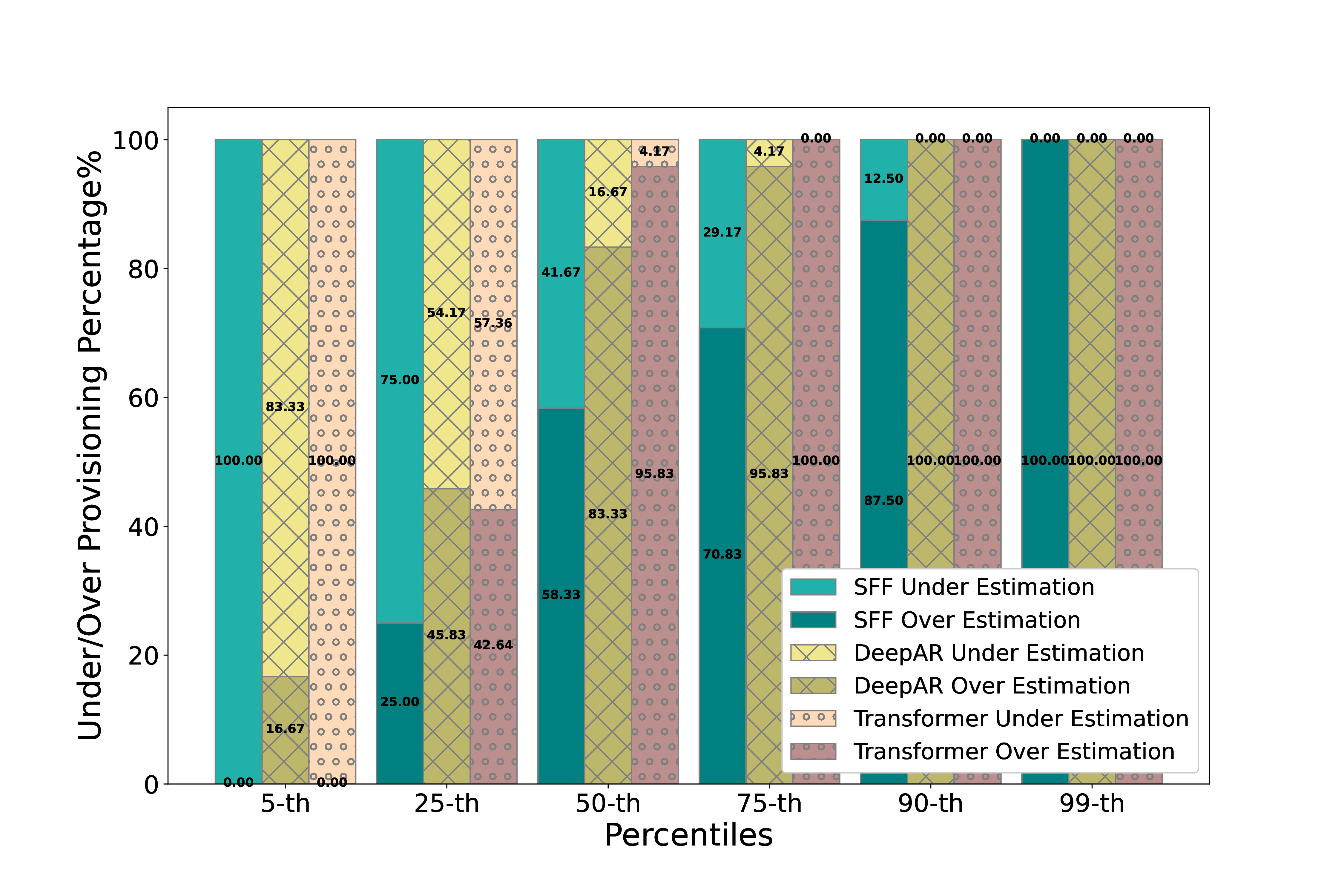}
\caption{Percentage of over/under-estimated PRBs.}
\label{Est_Plot}
\end{figure}

\subsection{PRB Provisioning Analysis}
First, we analyze the performance of the forecasting techniques in fulfilling the PRB allocation demand in the prediction range.
Fig. \ref{Est_Plot} shows the bar chart of the performance of the forecasting models, in terms of the percentage of overestimation and underestimation for different percentiles. Each bar is divided into two halves, with the upper and lower parts representing the percentage of underestimation and overestimation, respectively. The \ac{SFF} model has the highest underestimation, 100\%, when the forecast values of the 5-th percentile are taken into account, similarly, deepAR and Transformer also exhibit high under-estimation results. This shows that if we want to be conservative during predictions, SFF and Transformer would not fulfill the PRB allocation requirements in any case, which would compromise significantly the O-RAN performance. The results also show the trend of over-estimation increases with the percentile for all the methods. This can be seen as a major improvement because the allocated \ac{PRB}s are at least sufficient, but it will showcase a significant increase in power consumption due to this extra PRB allocation. Determining this for the different methods will be showcased in the next section.  
At extremely low percentiles capturing and learning the spatial patterns, trends and dependencies of the dataset during training becomes complex. Decisions about which estimators should be used and which percentile of forecasting is beneficial depend on the applications, their requirements, and their sensitivity to over- and underestimation.

\begin{table}[htp!]
\centering
\caption{Comparison of the performance of different forecast methods under different percentile values}
\begin{tabular}{|l|clllll|}
\hline
\multicolumn{1}{|c|}{\multirow{2}{*}{\textbf{\begin{tabular}[c]{@{}c@{}}Forecast \\ Models\end{tabular}}}} & \multicolumn{6}{c|}{\multirow{2}{*}{\textbf{}}} \\
\multicolumn{1}{|c|}{} & \multicolumn{6}{c|}{} \\ \hline
 & \multicolumn{6}{c|}{\textbf{MSE}} \\ \hline
\textbf{LSTM} & \multicolumn{6}{c|}{45.094} \\ \hline
\textbf{SFF} & \multicolumn{6}{c|}{49.147} \\ \hline
\textbf{DeepAR} & \multicolumn{6}{c|}{\textbf{0.370}} \\ \hline
\textbf{Transformer} & \multicolumn{6}{c|}{4.259} \\ \hline
 & \multicolumn{6}{c|}{\textbf{MAE}} \\ \hline
\textbf{LSTM} & \multicolumn{6}{c|}{5.146} \\ \hline
\textbf{SFF} & \multicolumn{6}{c|}{4.510} \\ \hline
\textbf{DeepAR} & \multicolumn{6}{c|}{\textbf{0.489}} \\ \hline
\textbf{Transformer} & \multicolumn{6}{c|}{1.849} \\ \hline
 & \multicolumn{6}{c|}{\textbf{MAPE(\%)}} \\ \hline
\textbf{LSTM} & \multicolumn{6}{c|}{19.13} \\ \hline
\textbf{SFF} & \multicolumn{6}{c|}{15.300} \\ \hline
\textbf{DeepAR} & \multicolumn{6}{c|}{\textbf{2.042}} \\ \hline
\textbf{Transformer} & \multicolumn{6}{c|}{8.43583} \\ \hline
 & \multicolumn{6}{c|}{\textbf{Normalized Deviation}} \\ \hline
\textbf{SFF} & \multicolumn{6}{c|}{0.077} \\ \hline
\textbf{DeepAR} & \multicolumn{6}{c|}{\textbf{0.008}} \\ \hline
\textbf{Transformer} & \multicolumn{6}{c|}{0.031} \\ \hline
 & \multicolumn{6}{c|}{\textbf{Percentiles}} \\ \hline
 & \multicolumn{1}{c|}{\textbf{5-th}} & \multicolumn{1}{c|}{\textbf{25-th}} & \multicolumn{1}{l|}{\textbf{50-th}} & \multicolumn{1}{c|}{\textbf{75-th}} & \multicolumn{1}{c|}{\textbf{90-th}} & \multicolumn{1}{c|}{\textbf{99-th}} \\ \hline
 & \multicolumn{6}{c|}{\textbf{Quantile Loss}} \\ \hline
\textbf{SFF} & \multicolumn{1}{l|}{8.22} & \multicolumn{1}{l|}{2.311} & \multicolumn{1}{l|}{0.811} & \multicolumn{1}{l|}{2.42} & \multicolumn{1}{l|}{5.66} & 11.39 \\ \hline
\textbf{DeepAR} & \multicolumn{1}{l|}{\textbf{0.746}} & \multicolumn{1}{l|}{\textbf{0.056}} & \multicolumn{1}{l|}{\textbf{0.434}} & \multicolumn{1}{l|}{\textbf{0.773}} & \multicolumn{1}{l|}{\textbf{1.210}} & \textbf{2.11} \\ \hline
\textbf{Transformer} & \multicolumn{1}{l|}{3.148} & \multicolumn{1}{l|}{0.218} & \multicolumn{1}{l|}{1.891} & \multicolumn{1}{l|}{4.171} & \multicolumn{1}{l|}{5.928} & 10.63 \\ \hline
 & \multicolumn{6}{c|}{\textbf{Coverage}} \\ \hline
\textbf{SFF} & \multicolumn{1}{l|}{0} & \multicolumn{1}{l|}{0.25} & \multicolumn{1}{l|}{0.583} & \multicolumn{1}{l|}{0.708} & \multicolumn{1}{l|}{0.875} & \textbf{1} \\ \hline
\textbf{DeepAR} & \multicolumn{1}{l|}{\textbf{0.166}} & \multicolumn{1}{l|}{\textbf{0.485}} & \multicolumn{1}{l|}{0.833} & \multicolumn{1}{l|}{0.958} & \multicolumn{1}{l|}{\textbf{1}} & \textbf{1} \\ \hline
\textbf{Transformer} & \multicolumn{1}{l|}{0} & \multicolumn{1}{l|}{0.458} & \multicolumn{1}{l|}{\textbf{0.958}} & \multicolumn{1}{l|}{\textbf{1}} & \multicolumn{1}{l|}{\textbf{1}} & \textbf{1} \\ \hline
\end{tabular}
\end{table}

To further quantify the forecasting performance, Table I compares the methods against a deterministic single point estimator \ac{LSTM} in terms of error metrics \cite{hochreiter1997long} like \ac{MSE}, \ac{MAE} and \ac{MAPE}. Since these metrics are specific to single-point estimates, we consider the median \ac{PRB}s values of predictions of probabilistic estimators to compute these metrics. It is essential to note that the performance of \ac{LSTM} in terms of all the error metrics is inferior compared to probabilistic estimators. Deterministic models are sensitive to outliers, uncertainties, and non-stationarity of the data, which makes them less effective and biased for forecasting. For comparison of performance of probabilistic estimator we consider metrics like \ac{ND}, \ac{QL}, and coverage \cite{salinas2020deepar}, \cite{Kasuluru_2023},\cite{EUCNC_2023} for different percentiles. DeepAR and Transformer show better results in all metrics, especially in the cases of \ac{ND}, \ac{QL}, and Coverage. This is because DeepAR models more accurately the forecasted distribution, promoting the generation of accurate forecasts with a reasonable degree of uncertainty. Transformers also use a self-attention mechanism and parallel processing of the input data. These two methods although more complex outperform the \ac{SFF}.
\begin{figure}[htp!]
\centering
\includegraphics[width=\linewidth]{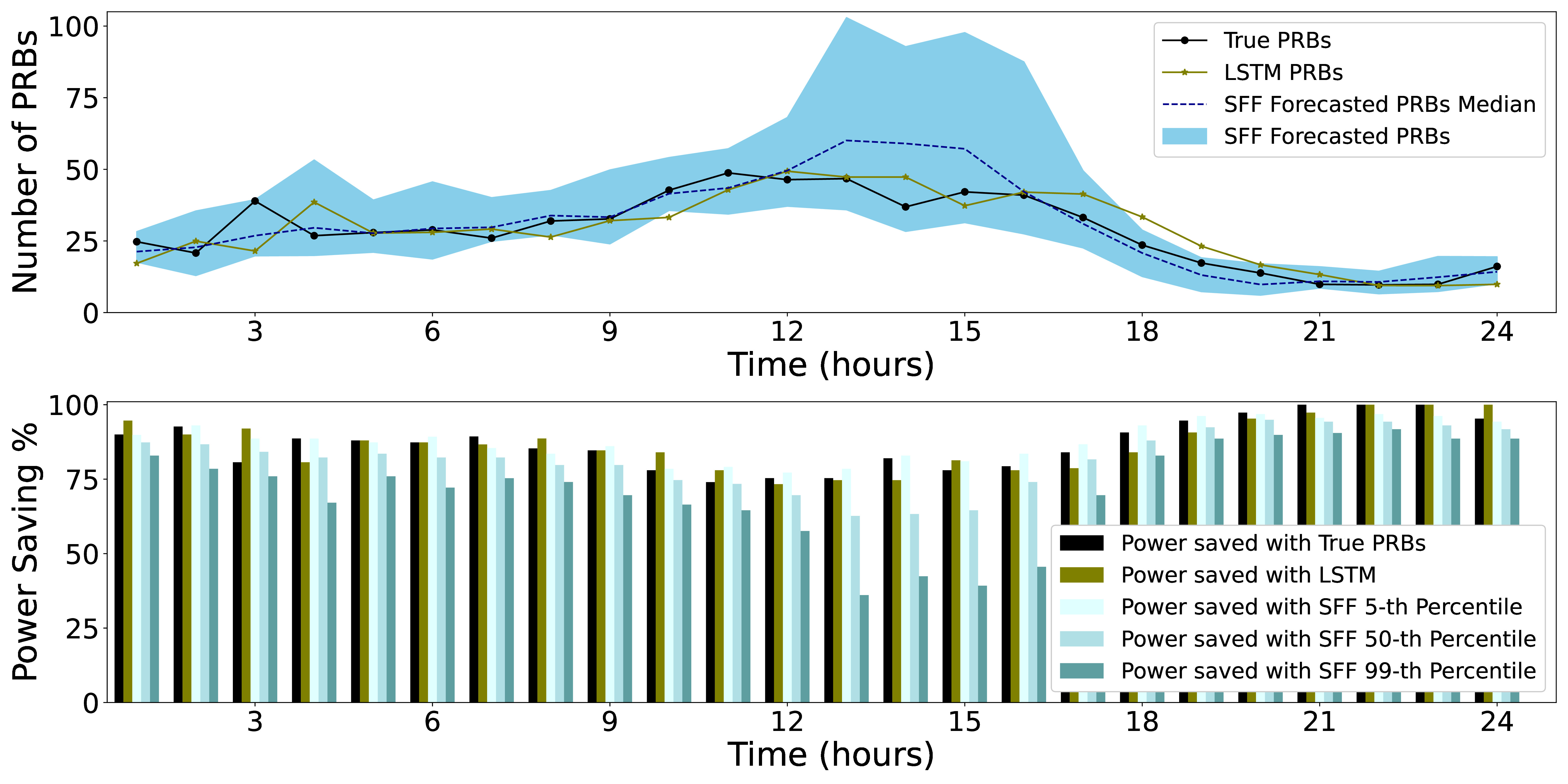}
\caption{Simple-Feed-Forward estimator with power saving.}
\label{E_Plot_sff1}
\end{figure}

\begin{figure}[htp!]
\centering
\includegraphics[width=\linewidth]{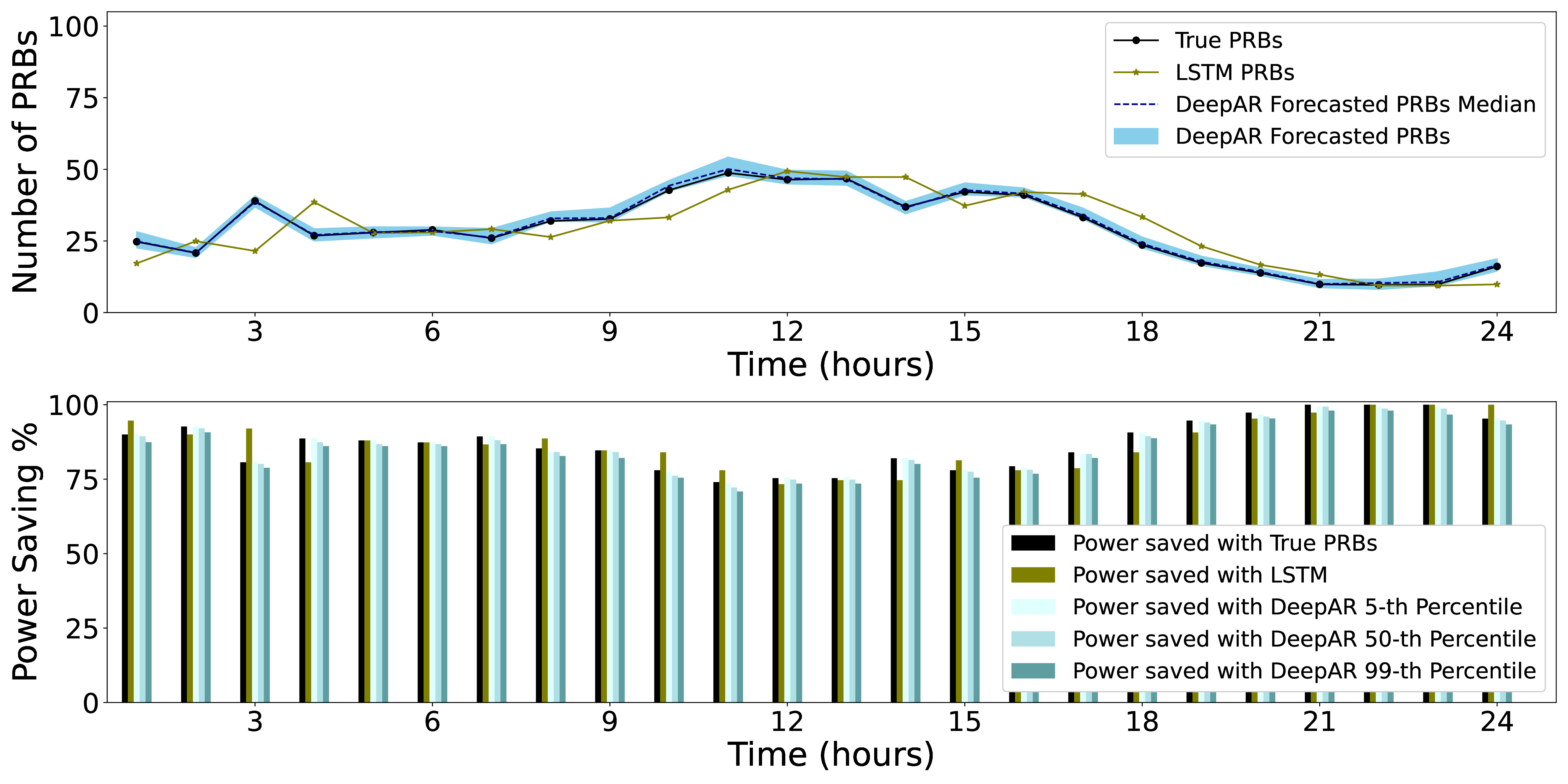}
\caption{DeepAR estimator with power saving.}
\label{E_Plot_da1}
\end{figure}

\begin{figure}[htp!]
\centering
\includegraphics[width=\linewidth]{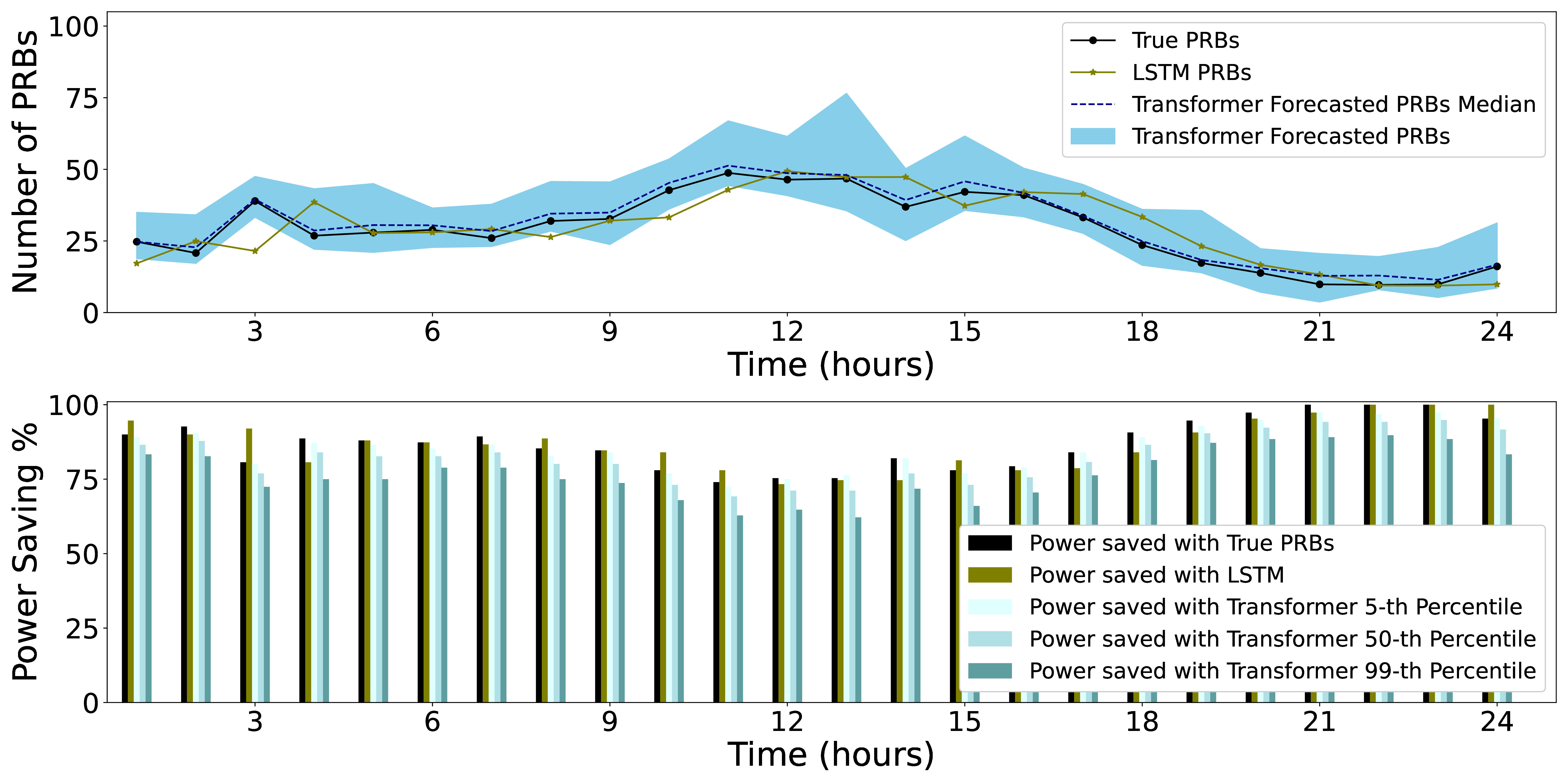}
\caption{Transformer estimator with power saving.}
\label{E_Plot_t1}
\end{figure}
\subsection{Sustainability Analysis}
As we have stated in Section II.B, there is a tradeoff between the PRB allocation and the total power consumption given by \eqref{eq:P_out}. For this analysis, Figures \ref{E_Plot_sff1}, \ref{E_Plot_da1} and \ref{E_Plot_t1} show the predictions of \ac{PRB}s using the different methods in the upper subplots along with their power saving every hour, in comparison with the maximum PRB usage (i.e., using the total amount of available PRBs 160). For the the lower subplot, the formulas in \eqref{eq:power} and \eqref{eq:P_out} have considered. In the upper graphs, the y-axis represents the number of \ac{PRB}s, and in the lower graph, the y-axis represents power saving as a percentage. In both subplots, the x-axis represents the length of the prediction data, i.e., 24 hours. The true \ac{PRB}s are shown in black, the \ac{LSTM} in green, the probabilistic estimators, and their median value in shades of blue in both subplots of all figures. For probabilistic methods, the area in shaded blue correspond to the 1-99 percentiles. In the lower subplots, the power saving in terms of \%, corresponding to the probabilistic estimators, are presented as a bar graph for only three percentiles, namely, the 5-th, 50-th, and 90-th percentiles, to show the impact of the predicted \ac{PRB}s at different percentiles on the power savings along with the bars for the True data and the \ac{LSTM} predicted data for comparison. From the above sections, it is known that the probabilistic estimators provide a spectrum or range of output predictions. This spectrum range is represented as percentiles ranging from the 1st to the 99th percentile.  In the upper subplot of Fig. \ref{E_Plot_sff1}, the blue shaded area shows the prediction of the True \ac{PRB}s by the \ac{SFF} estimator. From the large scatter of the \ac{PRB} prediction, it can be seen that the performance of the \ac{SFF} estimator is poor due to the large uncertainty. The fluctuation in power savings in the lower subplot also indicates that the \ac{SFF} is less consistent in predicting the data. It causes a high variation in power saving rate over 24 hours a day. 

DeepAR in Fig. \ref{E_Plot_da1} shows the best performance in terms of predictions with less uncertainty as the data spread is meager. When compared to the prediction of \ac{SFF} in Fig.\ref{E_Plot_sff1}, even transformers perform better prediction, as seen in Fig.\ref{E_Plot_t1}. 

As shown in Table I, \ac{LSTM} exhibits high error values, in terms of MSE, MAE and MAPE, because the single-point estimators are very sensitive to outliers and non-stationarity data. It also fails to recognize the complex patterns over a long data sequence. This fact corroborates the false sense of confidence in single-point forecasting techniques, which are not able to capture all the underlying uncertainties in their predictions. Furthermore, from the bottom plots of Fig.\ref{E_Plot_da1} and \ref{E_Plot_t1}, the DeepAR and transformer have consistent power savings throughout the day. It should be noted that a increase in the percentile reduces the power saving, because it results in higher PRB allocation, and this fact has a consequence an increase in the total consumed power. This is further clarified in Table II that shows the trade-off between power saving and PRB over/under-provisioning.

\begin{table*}[htp!]
\centering
\caption{Comparison of power saving, and Over-/Under provisioning \% of forecast methods under different percentile values}
\begin{tabular}{|lcccccc|}
\hline
\multicolumn{7}{|c|}{\textbf{\begin{tabular}[c]{@{}c@{}}Baseline\\ Models\end{tabular}}} \\ \hline
\multicolumn{1}{|c|}{\textbf{True Data}} & \multicolumn{3}{c|}{\textbf{Power Saving\%}} & \multicolumn{3}{c|}{87.1} \\ \hline
\multicolumn{1}{|c|}{\multirow{3}{*}{\textbf{LSTM}}} & \multicolumn{3}{c|}{\textbf{PowerSaving\%}} & \multicolumn{3}{c|}{86.7} \\ \cline{2-7} 
\multicolumn{1}{|c|}{} & \multicolumn{3}{c|}{\textbf{Overprovisioning\%}} & \multicolumn{3}{c|}{50} \\ \cline{2-7} 
\multicolumn{1}{|c|}{} & \multicolumn{3}{c|}{\textbf{Underprovisioning\%}} & \multicolumn{3}{c|}{50} \\ \hline
\multicolumn{1}{|c|}{\multirow{2}{*}{\textbf{\begin{tabular}[c]{@{}c@{}}Forecast\\ Models\end{tabular}}}} & \multicolumn{6}{c|}{\textbf{Percentiles}} \\ \cline{2-7} 
\multicolumn{1}{|c|}{} & \multicolumn{1}{c|}{\textbf{5-th}} & \multicolumn{1}{c|}{\textbf{25-th}} & \multicolumn{1}{l|}{\textbf{50-th}} & \multicolumn{1}{c|}{\textbf{75-th}} & \multicolumn{1}{c|}{\textbf{90-th}} & \textbf{99-th} \\ \hline
\multicolumn{1}{|l|}{} & \multicolumn{6}{c|}{\textbf{Power Saving \%}} \\ \hline
\multicolumn{1}{|l|}{\textbf{SFF}} & \multicolumn{1}{c|}{87.7} & \multicolumn{1}{c|}{83.78} & \multicolumn{1}{c|}{81.69} & \multicolumn{1}{c|}{79.5} & \multicolumn{1}{c|}{77.2} & 71.4 \\ \hline
\multicolumn{1}{|l|}{\textbf{DeepAR}} & \multicolumn{1}{c|}{86.9} & \multicolumn{1}{c|}{86.4} & \multicolumn{1}{c|}{86.1} & \multicolumn{1}{c|}{85.8} & \multicolumn{1}{c|}{85.5} & 84.9 \\ \hline
\multicolumn{1}{|l|}{\textbf{Transformer}} & \multicolumn{1}{c|}{85.7} & \multicolumn{1}{c|}{83.6} & \multicolumn{1}{c|}{82.5} & \multicolumn{1}{c|}{81.2} & \multicolumn{1}{c|}{79.8} & 76.8 \\ \hline
\multicolumn{1}{|l|}{} & \multicolumn{6}{c|}{\textbf{Overprovisioning \%}} \\ \hline
\multicolumn{1}{|l|}{\textbf{SFF}} & \multicolumn{1}{c|}{0} & \multicolumn{1}{c|}{25} & \multicolumn{1}{c|}{58.3} & \multicolumn{1}{c|}{70.83} & \multicolumn{1}{c|}{87.5} & 100 \\ \hline
\multicolumn{1}{|l|}{\textbf{DeepAR}} & \multicolumn{1}{c|}{16.6} & \multicolumn{1}{c|}{45.83} & \multicolumn{1}{c|}{83.33} & \multicolumn{1}{c|}{95.83} & \multicolumn{1}{c|}{100} & 100 \\ \hline
\multicolumn{1}{|l|}{\textbf{Transformer}} & \multicolumn{1}{c|}{0} & \multicolumn{1}{c|}{45.833} & \multicolumn{1}{c|}{95.83} & \multicolumn{1}{c|}{100} & \multicolumn{1}{c|}{100} & 100 \\ \hline
\multicolumn{1}{|l|}{} & \multicolumn{6}{c|}{\textbf{Underprovisioning \%}} \\ \hline
\multicolumn{1}{|l|}{\textbf{SFF}} & \multicolumn{1}{c|}{100} & \multicolumn{1}{c|}{75} & \multicolumn{1}{c|}{41.66} & \multicolumn{1}{c|}{29.16} & \multicolumn{1}{c|}{12.5} & 0 \\ \hline
\multicolumn{1}{|l|}{\textbf{DeepAR}} & \multicolumn{1}{c|}{83.334} & \multicolumn{1}{c|}{54.1} & \multicolumn{1}{c|}{16.6} & \multicolumn{1}{c|}{4.16} & \multicolumn{1}{c|}{0} & 0 \\ \hline
\multicolumn{1}{|l|}{\textbf{Transformer}} & \multicolumn{1}{c|}{100} & \multicolumn{1}{c|}{54.16} & \multicolumn{1}{c|}{4.16} & \multicolumn{1}{c|}{0} & \multicolumn{1}{c|}{0} & 0 \\ \hline
\end{tabular}
\end{table*}

Table II compares the performance of \ac{SFF}, DeepAR, and Transformer forecasting models with the baseline model \ac{LSTM} in terms of power savings, overestimation, and underestimation in percentage. It helps to analyze and understand the trade-off between under/over-provisioning and power saving. Ground truth baseline, denoted in the table as True Data, shows the highest potential of power saving without compromising the network performance (i.e., without suffering under-provisioning), around 87.1\%. Even though the \ac{LSTM} performance degradation in terms of power saving is small, it is poorly performing in terms of PRB allocation, providing an over/under-provision rate of 50\%.

In the probabilistic models in Table II, Table I, and the figures, it is clear that reducing the percentile increases the power saving, but at the expense of under-provisioning. However, when over-provisioning network performance is guaranteed, at expenses of a higher power consumption. The critical point is to quantify the trade-off, i.e, how much over-provisioning is admissible so that the impact of power saving is not significant. Looking at Table II, for the 99-th percentile, DeepAR exhibits a 84.9\% power saving. This confidence interval provides a power saving similar to the ground truth (i.e, true data) 87.1\%. In general, overestimation leads to wasted resources, high operating costs, and reduced power efficiency, while underestimation can lead to unmet demand, bottlenecks, and operational disruptions. 

Compared to the other estimators in Table II, \ac{SFF} suffers greatly from over-/under provisioning. Even with \ac{PRB}s at the 90-th percentile, there are significant prediction deviations. The performance of DeepAR and Transformer is balanced due to the lower uncertainty in the predictions of \ac{PRB} shown in Fig. 4 and 5, and the better understanding of data trends and seasonality. To summarize, each model is unique in terms of its strengths and weaknesses in resource prediction and energy savings. Thus, which percentile of values should be selected for resource allocation depends on the application, available power, and sensitivity to over/under-estimation of resources.

\section{Conclusions and Future Directions}
\label{conclusions}

In this paper, we have proposed a novel approach to characterize the \ac{PRB} load in sustainable \ac{O-RAN} using probabilistic forecasting techniques. We first provided background information on the O-RAN architecture and components, emphasizing the importance of power consumption models for sustainable implementations.  We then discussed probabilistic forecasting techniques, including \ac{SFF} and DeepAR, and evaluated their effectiveness in characterizing \ac{PRB} load dynamics based on simulation results. The results show the potential of probabilistic forecasting to improve resource management and energy efficiency in \ac{O-RAN} deployments. In particular, DeepAR estimators are shown to predict \ac{PRB}s with lower uncertainty and effectively capture the temporal dependencies in the dataset compared to \ac{SFF}- and Transformer-based models. Different percentile selections for probabilistic methods can also increase power savings, but at the cost of over/under-provisioning.  At the same time, the performance of the \ac{LSTM} is shown to be inferior to the probabilistic estimators in terms of all error metrics. For the future, there are several opportunities for further research, such as integration with energy optimization techniques (including dynamic power management strategies or renewable energy integration), validation in the real world with field trials and pilot studies, and exploration of new architectures together with additional data sources to improve accuracy.




\balance

\bibliographystyle{ieeetr}
\bibliography{biblio}  

\end{document}